\documentclass[aps,prb,twocolumn,amsmath,amssymb,superscriptaddress,floatfix,showpacs]{revtex4}

\usepackage[dvips]{graphicx}

\begin{document}
\title{Cyclic exchange, isolated states and spinon deconfinement 
in an XXZ Heisenberg model on the checkerboard lattice}

\author{Nic Shannon}
\affiliation{
Department of Advanced Materials Science,
Graduate School of Frontier Sciences, University of Tokyo, 5--1--5, 
Kashiwahnoha, Kashiwa, Chiba 277--8851, Japan}
\affiliation{
CREST, Japan Science and Technology Agency, Kawaguchi 332-0012, Japan
}

\author{Gr\'egoire Misguich}
\affiliation{
Service de Physique Th\'eorique,
CEA-Saclay, 91191 Gif-sur-Yvette C\'edex, France}

\author{Karlo Penc}
\affiliation{
Research Institute  for  Theoretical Solid State  Physics   and
Optics, H-1525 Budapest, P.O.B.  49, Hungary}

\date{\today}

\begin{abstract}
The antiferromagnetic Ising model on a checkerboard lattice has an ice--like
ground state manifold with extensive degeneracy.  and, to leading 
order in $J_{xy}$, deconfined spinon excitations.  We explore the role 
of cyclic exchange arising at order $J^2_{xy}/J_z$ on the ice states and 
their associated spinon excitations.   By mapping the original problem
onto an equivalent quantum six--vertex model, we identify three different 
phases as a function of the chemical potential
for flippable plaquettes  --- a phase with long range N\'eel order and 
confined spinon excitations, a non--magnetic state of resonating square 
plaquettes, and a quasi-collinear phase with gapped but deconfined spinon excitations.  
The relevance of the results to the 
square--lattice quantum dimer model is also discussed.
\end{abstract}


\maketitle

The past decade has seen a great renaissance in the study of frustrated 
quantum spin systems.  On the experimental front, advances in the synthesis of 
magnetic oxides have given rise to a great wealth of new frustrated materials with 
highly unusual and interesting properties.   And, at the same time, highly frustrated 
models have become a favorite playground of theorists 
seeking to understand unconventional phase transitions and excitations.

\begin{figure}[b]
  \centering
  \includegraphics[height=4.0truecm]{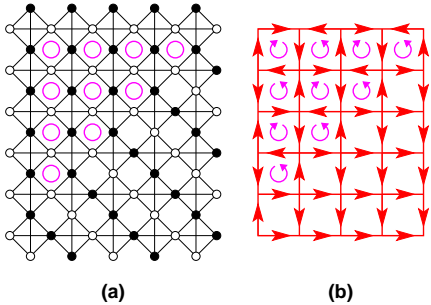}
\caption{
a) The checkerboard lattice on which the ice states,
 and 
b) the square lattice on which the states of the 
six--vertex model are defined.
Any Ising state obeying the ice rules, 
e.g. that shown in (a), is equivalent to (b) six--vertex model configuration. 
In the state shown, the upper left corner has N\'eel order, while
the lower right corner has collinear order. Flippable plaquettes
are denoted with circles.  In the case of the six--vertex model,
these have a definite sense of rotation
\label{fig:eq}}
\end{figure}

Recently, it was proposed that the geometric
frustration present on the pyrochlore lattice could give rise to
fractional charges in two or three dimensions, \cite{fulde} in a 
physically realistic model based on strong nearest neighbor repulsion 
close to commensurate filling.\cite{anderson}  
The charge ordering problem considered in [\onlinecite{fulde}] 
is classically equivalent to one of Ising antiferromagnetism, 
and in this paper we 
consider the simplest  possible test case for these ideas, the 
XXZ Heisenberg model on a checkerboard (2D pyrochlore) lattice.
We proceed by mapping this model onto an equivalent, {\it quantum
six--vertex model} (Q6VM), and describe the nature of the 
ground state and low lying spin excitations of this model
as a function of a control parameter $V$, which acts as a chemical 
potential  for those ``flippable plaquettes'' accessible  to cyclic
exchange.

We identify three different ground states, a phase with long range N\'eel order,  
a non--magnetic state of resonating square plaquettes, and a 
partially disordered phase of ``isolated states'' 
with extremely large ground state degeneracy, referred to as the 
``quasi--collinear'' phase below.
  Because of the anisotropy of the model, 
all spin excitations are gapped.  
It is possible to identify the lowest lying excitations of the N\'eel phase
as spin waves, and those of the quasi--collinear phase as deconfined spinons.
We also identify the special role of the isolated states in supporting fractional excitations.
Many of these results are also relevant to the much studied square 
lattice quantum dimer model (QDM).\cite{rk88}

{\it Model and mapping onto Q6VM.}
We take as a starting point the spin--1/2 
anisotropic Heisenberg model with antiferromagnetic 
interactions, $J_z, J_{xy} >0$, in the limit  $J_z \gg J_{xy}$
  \begin{equation}
      \label{eqn:H}
    \mathcal{H} = J_z \sum_{\langle ij \rangle} S_i^z S_j^z + 
    \frac{J_{xy}}{2} \sum_{\langle ij \rangle} 
     \left(S_i^+ S_j^- +S_i^- S_j^+\right)  \;.
  \end{equation}
Here the sum $\sum_{\langle ij \rangle}$ runs over the bonds 
of the 2D pyrochlore or checkerboard lattice, shown in 
Fig.~\ref{fig:eq}a).
In the Ising limit, 
$J_{xy} =  0$, this model has an extensive 
ground state degeneracy  --- every state with exactly two up and 
two down spins per tetrahedron (cross linked square) is a ground state.
For historical reasons, this is known as the ``ice rules'' constraint.
Topologically, ``ice'' states have the structure of closely 
packed loops of up and down spins, and are separated by
a gap $J_z$ from the lowest lying excited state.
Flipping any given down spin connects two adjacent 
loops of up spins, creating two ``T--junction'' like topological defects 
(spinons), which propagate independently.\cite{fulde,hfb03}  
The pyrochlore (checkerboard) lattice is  bipartite in tetrahedra.  
Spinons are created in A and B sublattice pairs, and move so as to preserve 
tetrahedron sublattice.

By drawing an arrow from the center of A to B sublattice tetrahedra where they 
share an up spin, and from B to A where they share a down spin, 
one can show that the many ground states of the Ising model on a checkerboard lattice are
in exact, one--to--one correspondence with the states of the 
classical {\it six vertex model} (6VM),\cite{kohanoff,mts01} widely studied as a 2D analogue of
water ice.   From this mapping, we know that 
a) the ground state manifold of the Ising model grows as 
$ W \propto (4/3)^{3N/4}$
where $N$ is the number lattice sites \cite{lieb}
and b) all correlation functions decay algebraically.\cite{baxter} 

\begin{figure}[b]
  \centering
  \includegraphics[width=7.5truecm]{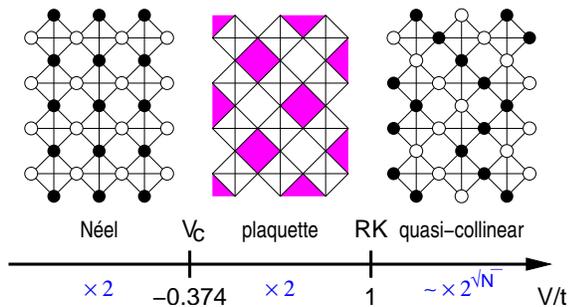}
\caption{
The phase diagram of the model as a function of $V/t$. 
The N\'eel phase breaks the point group, while the plaquette phase 
breaks the
translational symmetry.
The Rokhsar--Kivelson point is marked RK.
\label{fig:pd}}
\end{figure}

Up to this point, our analysis contains only classical statistical 
mechanics and simple topological arguments. 
Quantum mechanics reenters the problem when we consider a small but 
finite $J_{xy}\ll J_z$.
In this case, the ice states are no longer eigenstates.  Short lived 
virtual excitations enable the system to tunnel between different ice 
state configurations wherever pairs of up spins and down spins
occur diagonally opposite one another on one of the empty 
square plaquettes of the checkerboard lattice.\cite{footnote1}
The allowed reconfigurations of these ``flippable plaquettes''
can be described within 
degenerate perturbation theory by the effective Hamiltonian
\begin{equation}
    \label{eqn:Heff}
 \mathcal{H}_{\rm 2nd} = - \frac{J_{xy}^2}{J_z} 
  \sum_{\square} (
   S^+_1 S^-_2 S^+_3 S^-_4+ 
   S^-_1 S^+_2 S^-_3 S^+_4) \;,
\end{equation}
where the indices $1$ to $4$ count consecutive sites 
(either clockwise or anticlockwise), of an empty plaquette.\cite{mts01}

In terms of the 6VM representation, Eq.~(\ref{eqn:Heff})
acts on a plaquette where four arrow are joined nose to tail, so
as to invert all of the arrows and change the sense of rotation of 
the plaquette (c.f.~[\onlinecite{chakravarty02}]).
%The quantum dynamics in the Q6VM we consider are directly analogous 
%to the resonance of dimers in the QDM considered by 
%Rokhsar and Kivelson (RK),\cite{rk88} as an simplified model of a short ranged
%resonating valence bond state.\cite{And73}
The quantum dynamics in the Q6VM we consider are directly analogous 
to the resonance of dimers in the QDM,\cite{rk88} studied as an simplified 
model of a 
%short ranged 
resonating valence bond state.\cite{And73}
Formally, in fact, the Hamiltonian is exactly the same, although the 
Hilbert space on which it acts is different.
And, as in the QDM, we anticipate that quantum effects will in general 
select a ground state with finite degeneracy from the vast manifold of 
classically--allowed ice states.  

As such, there is only one (kinetic) energy scale in the problem, 
$t = J_{xy}^2/J^z$. However in order to study the different possible 
phases of the model it is useful to introduce a further control parameter.  
A suitable control parameter for the QDM is a diagonal term which counts 
the number of dimers which can resonate in any given dimer covering.
By direct analogy, we introduce a diagonal interaction $V$ to the Q6VM which 
counts the number of flippable plaquettes
\begin{equation}
  \mathcal{H} =  \sum_\square \left[
 V \bigl(
 \mid \circlearrowleft \rangle\langle \circlearrowleft \mid \!+\! 
 \mid \circlearrowright \rangle\langle \circlearrowright \mid \bigr) 
 -t \bigl(
\mid \circlearrowleft \rangle\langle \circlearrowright \mid \!+\! 
\mid \circlearrowright \rangle\langle \circlearrowleft \mid \bigr)\right]
\;,
\label{eq:Q6VH}
\end{equation}
where the $\mid \circlearrowleft \rangle$ and $\mid \circlearrowright \rangle$
states represent  squares with the respective circular arrow configuration 
on the square edges, as seen in Fig.~\ref{fig:eq}(b).  We note that,
for a system with periodic boundary conditions, 
the net flux of vertex arrows through any given horizontal or vertical
cut defines a set of winding numbers which are
conserved by the Hamiltonian~(\ref{eq:Q6VH}) .

Our approach to determining the different phases of the 
Hamiltonian~(\ref{eq:Q6VH}) is the numerical diagonalization of
clusters with periodic boundary conditions of up to 64 spins, 
within the ice rules manifold of states,
supplemented with topological and symmetry arguments.
Details of these, together with further analysis of the related 
fermionic charge--ordering problem will be discussed further in separate 
publications.\cite{karlo+nic,erich+fulde}

\begin{figure}[bt]
  \centering
  \includegraphics[width=7.truecm]{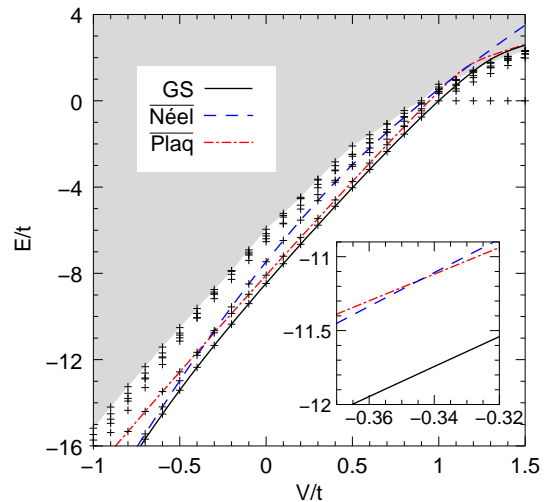}
\caption{
Energy level diagram of the 32--site pyrochlore--slab with periodic boundary
conditions as a function of
$V/t$, obtained by numerical diagonalization.  
We have shown the first 8 levels. Inset: the first two excited states cross 
at $V_c/t = -0.3437$ (the axes are the same as of the main plot).
\label{fig:el}}
\end{figure}

{\it Phase diagram.}
We first consider the nature of the ground state as a function the 
chemical potential for flippable plaquettes, $V$.
Our results are summarized in the phase diagram Fig.~\ref{fig:pd},
and the numerical evidence for each phase discussed below.

\begin{figure}[t]
  \centering
  \includegraphics[width=6.5truecm]{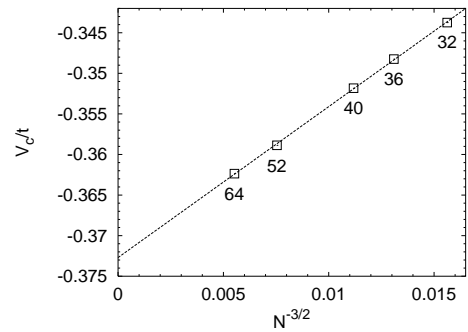}
\caption{
Estimate of the phase boundary between the N\'eel and plaquette
phases.  Empirically, the values of $V_c$ where the level crossings occur 
scale as $-0.3727 + 1.86 N^{-3/2}$.   Values are shown
for 32, 36, 40, 52, and 64 pyrochlore--slab sites. 
\label{fig:vc}}
\end{figure}

Negative values of $V$ favor states with flippable plaquettes.
The state with the greatest possible number of flippable plaquettes
is the N\'eel state, and this must be the grounds state for $V \to -\infty$.
The N\'eel state is two fold degenerate in the thermodynamic limit.   
For finite $V/t$, in a finite size system, quantum fluctuations lift this 
degeneracy, as seen in the low-energy spectrum of the Q6VM (Fig.~\ref{fig:el}).
We find a single phase for $V \lesssim -0.3t$, which we identify as the 
N\'eel phase.
Both the the symmetric and antisymmetric combinations of the
two symmetry-breaking  N\'eel ground states are visible in the spectrum, 
marked ``GS'' 
and ``$\overline{Neel}$''  respectively.   At a value of $V \sim -0.3t$, 
a third energy level, marked ``$\overline{Plaq}$'' 
crosses the first excitation ``$\overline{Neel}$'' .
We interpret this as evidence for a quantum phase transition
into a resonating plaquette phase, discussed below.
 From finite size scaling of the spectrum (Fig.~\ref{fig:vc})
 we estimate the critical value to be $V_c = -0.3727t$
in the thermodynamic limit.
As the competing N\'eel and plaquette order parameters break
lattice symmetries in very different ways, the transition between 
them is presumably of first order.

We find a single phase extending from $ -0.3t\lesssim V \leq t$,
including the XXZ point $V=0$.  
This phase terminates in the special high symmetry point $V=t$ for 
which the Hamiltonian (\ref{eq:Q6VH}) of the Q6VM
can be written as a sum of projection operators :
\begin{equation}
    \label{eq:RK}
  \mathcal{H_{\rm RK}} =  t \sum_\square \bigl(
\mid \circlearrowleft \rangle - \mid \circlearrowright \rangle \bigr)
 \bigl( \langle \circlearrowleft \mid - \langle \circlearrowright 
 \mid \bigr) \;.
\end{equation}
Following Rokhsar and Kivelson (RK),\cite{rk88} we can construct a zero 
eigenvalue state
of the $\mathcal{H_{\rm RK}}$ by taking the linear combination of
all the states in a given topological sector with the same amplitude.
Since this state is annihilated by the positive semi--definite
$\mathcal{H_{\rm RK}}$,  it must be a ground state.  
As in the QDM, static correlations can be computed exactly at this point.
Like the correlation functions of the 6VM, they decay algebraically with 
distance.

At the RK point, kinetic and potential energy are perfectly balanced; 
in the plaquette phase kinetic energy dominates, and resonating 
plaquettes repel one another so as gain the maximum kinetic 
energy.\cite{moessner04} 
The resulting state is 
essentially a Peierls--like distortion of the 
RK state in which only A (B) sublattice plaquettes resonate
$ \prod_{\square_{A(B)}}(\mid\circlearrowleft \rangle + \mid \circlearrowright \rangle)$.
The way in which the phase breaks lattice symmetries ---
it is two--fold degenerate, and invariant under operations
which map the alternating A and B plaquette sublattice onto themselves --- 
suggest the plaquette phase of the Q6VM is an Ising analogue of the a 
SU(2) valence--bond crystal of resonating plaquettes.
Such a phase has been proposed in the context of the square lattice 
QDM.\cite{lcr96}
Furthermore, the ground state of the 
Heisenberg--model on a checkerboard lattice is a valence bond crystal
of SU(2) singlets formed on alternate empty plaquettes,\cite{sindzingre2002}
with a possibility of an adiabatic continuity
between the ground state of the XXZ and SU(2) symmetric Heisenberg
models.\cite{philippeunpub}

For $V>t$ the ground state is the 
highly degenerate manifold of ``isolated'' states with no flippable 
plaquettes. 
They are eigenstates with 0 energy for any 
value of $V/t$, and become the ground state for $V>t$.\cite{suto}
The prototype of an isolated state is the collinear configuration shown in 
Fig~\ref{fig:eq}.  In this reference state all vertex arrows point from 
left to right or from top to bottom.
Inverting the direction of the arrows along an 
arbitrary number of lines, subject to the constraint that all of them 
are either horizontal or vertical, creates new isolated states. 
This leads to a ground 
state degeneracy which grows as $4 (2^p - 1)$ for regularly shaped clusters, 
where $p \sim \sqrt{N}$.
In these states, the direction of arrows along either the horizontal or 
vertical lines is long-range ordered, but quantum effects none the 
less fail to select a ground state with finite degeneracy. 
We refer to this phase of the Q6VM as ``quasi--collinear''. 
Finally, since the transition between the quasi--collinear
 phase and the resonating
plaquette phase takes place through the softening of specific 
excitation (discussed below), we identify it as second order.

\begin{figure}[t]
    \centering
  \includegraphics[width=7.5truecm]{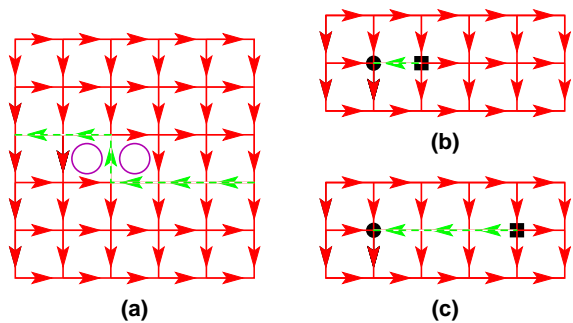}
 \caption{(a) A ``leap--frog'' excitation in the quasi--collinear phase.  Two flippable
   plaquettes (denoted with circles) are created by reversing the arrows 
   of the collinear reference state on a line with a single one--step kink.  
   The motion of the pair of flippable plaquettes is equivalent to a
   one--dimensional hopping model with an energy spectrum 
   $\varepsilon(k)=2V + 2t\cos k$, where $k$ is  an effective one--dimensional momentum.  
   (b) and (c):
   The deconfined spinons in the collinear phase (black dots). 
   Note that spinons hop so as to stay within a given 
  sublattice of the (bi--partite) square lattice.
 \label{fig:1dx}}
\end{figure}

{\it Excitations.}
First let us first consider the nature of excitations at fixed $S^{z} = 0$.
A state with $n$ flippable plaquettes has a diagonal matrix element
$nV$ and is connected to $n$ other states. Gerschgorin's theorem 
places a bound $|H_{ii}-\varepsilon_i|<\sum_{j} |H_{ij}|$ on the separation of the 
$i$-th eigenvalue $\varepsilon_i$ from the diagonal matrix element $H_{ii}$. 
In the case in point, this bound is $|nV-\varepsilon_i|< n t$, or 
$n(V-t)<\varepsilon_i< n(V+ t)$. The smallest energy in an arbitrary
topological sector is thus larger than $V-t$, which gives a lower bound on
the value of the gap in the quasi--collinear phase for $V>t$. 
This above argument permits a gapless spectrum at the RK 
point $V=t$.  In fact it is possible to explicitly construct a 
family of states with a gap that vanishes at the RK point, as
shown in Fig.~\ref{fig:1dx}(a):  
the energy spectrum of this particular excitation forms a continuum between 
$2V-2 t$ and $2V+2t$.

Now let us consider spin excitations with $S^{z}= \pm 1$.  
If we neglect virtual processes at order $J_{xy}^{2}/J_{z}$, and the 
possibility of entropic confinement at finite temperature,
these propagate as independent fractional excitations.\cite{fulde}   
Quantum effects may, or may not, act to confine these excitations, 
depending on the type of correlations present in the ground state they select.  
The N\'eel ground state has a two fold ground state degeneracy, and 
separating the topological defects created by flipping a spin creates a string of 
unflippable plaquettes.  This leads to confinement
of spinons, and the low lying spin excitations of the N\'eel phase of
our model have the same quantum numbers as a spin wave.
On general grounds, we expect the same to be true of the
plaquette phase.

The manifold of isolated states selected by $V$
 {\it can} support deconfined spinons, however.
Since no new flippable plaquettes are introduced into isolated states 
by flipping a single spin, and the pair of topological defects created by 
flipping a single spin can be separated without
creating new flippable plaquettes, spinons are deconfined.  
An example of a pair of deconfined spinon excitations is 
shown in Fig.~\ref{fig:1dx}(b)-(c).   For $V \gg J_{xy} \gg t$, spinon 
motion is movement is confined to the x and y directions, 
but by scattering off one another, a pair of spinons can explore 
the full two dimensional space of the lattice.  Whether  a more
general class of deconfined spinon excitation becomes possible 
as one approaches the RK point remains an open question.
We also note that while fixing the boundary conditions will lift the 
degeneracy of the isolated state manifold, it need not affect 
the arguments for spinon deconfinement presented above.

{\it Conslusions}
We have established that, as a function of the chemical potential for 
``flippable plaquettes'' accessible to cyclic exchange, the XXZ Heisenberg model 
on a checkerboard lattice exhibits N\'eel, resonating plaquette and 
quasi--collinear phases.   If virtual processes at $J_{xy}^2/J_z$ are 
ignored, spinon excitations in the XXZ Heisenberg model 
are deconfined.   We have shown explicitly that a subset of spinon excitations  
--- those associated with isolated states --- remain deconfined even when these 
quantum effects are taken into account.  
Finally, we mention that the equivalents of both the ``leap-frog'' and spinon 
excitations can also be constructed in the square--lattice QDM 
for $V>t$.\cite{footnote2} 

We are pleased to acknowledge helpful discussions with 
P. Fazekas, P. Fulde, R. Moessner, 
V.  Pasquier, E. Runge, D. Serban, A. S\"ut\H o, M.~Roger,
P. Sindzingre and P. Wiegmann. We thank the support of the
Hungarian OTKA T038162 and T037451,
EU RTN ``He~III Neutrons'' and the 
guest program of MPI-PKS Dresden.

\end{document}